\documentclass[fleqn,twoside]{article}
\usepackage{espcrc2}

\usepackage{graphicx}
\usepackage[figuresright]{rotating}

\newcommand{\AmS}{{\protect\the\textfont2
 A\kern-.1667em\lower.5ex\hbox{M}\kern-.125emS}}

\title{Exclusive Charmless $B$ Decays in QCD \hspace{4cm}{\bf \small SI-HEP-2006-08} }

\author{Alexander Khodjamirian~\address[MCSD]{~Theoretische Physik 1, Fachbereich Physik,\\
Universit\"at Siegen, D-57068 Siegen, Germany 
        }%
\thanks{~Invited talk at the First Workshop on Theory, Phenomenology and Experiments in Heavy Flavour Physics, 
Anacapri, Italy, May 29-31 2006.}
}
\begin{document}
\begin{abstract}
The problem of hadronic input
in charmless nonleptonic $B$ decays is discussed. QCD sum rules
and their light-cone versions (LCSR) provide an important part of 
this input, such as the decay constant $f_B$ and $B\to \pi$ form factor. 
Employing the LCSR technique, the $B\to \pi\pi$ hadronic matrix elements 
with emission, penguin and  annihilation topologies 
are calculated, with  no evidence for large nonfactorizable effects and/or 
strong phases.

\vspace{1pc}
\end{abstract}

\maketitle

\section{INTRODUCTION}

Charmless hadronic $B$ decays proceed due to a  
unique interference of electroweak and strong interactions, allowing
one to observe CP-violation effects within and beyond Standard Model (discussed in the talks by M.~Gronau and R.~Fleischer). 
On the experimental side, data  on  branching fractions and CP-asymmetries are being steadily accumulated \cite{HFAG} in many 
$B\to h_1 h_2$ channels ($h_{1,2}=\pi,K,\rho,K^*$ and other light mesons). 
Still there are noticeable  differences between BaBar  and Belle   
results. On the theory side, there is a big challenge  
of calculating the relevant hadronic input in QCD, 
which is the main topic of my talk. 

\section{THEORY INPUTS IN CHARMLESS \\$B$ DECAYS}

Integrating out the electroweak 
and quark-gluon interactions at short distances  and constructing
the effective weak Hamiltonian, one arrives at a 
generic expression for the charmless decay amplitude, e.g., for 
$ \bar{B}^0 \to \pi^+\pi^-$:
\begin{eqnarray}
A(\bar{B}^0 \to \pi^+\pi^-) = 
\langle \pi^+ \pi^- |H_{eff}|\bar{B}^0 \rangle
\nonumber\\
=\frac{G_F}{\sqrt{2}}\!\sum_{T=E,P_c,A,..}
\!\sum_i  \lambda_i c_i(\mu)\langle \pi^+ \pi^- |O_i|\bar{B}^0 \rangle^{T}_{\mu}\,,
\label{ampl}
\end{eqnarray}
where $\lambda_i$ and $c_i$ are
the CKM factors and Wilson coefficients, respectively,  
and $\mu\sim m_b$ is the renormalization
scale for $H_{eff}$. Each operator $O_i$ in the effective 
Hamiltonian generates several hadronic matrix elements 
with different {\em topologies} $T$, that is, the ways
of contracting the quark fields in the initial $B$ meson
and final pions. Here, $T=E,P_c,A,..,$ denotes
emission, charming penguin, annihilation, etc. 
Similar to Eq.(\ref{ampl}), 
the hadronic input for each nonleptonic channel $B\to h_1 h_2$ 
is encoded in a set of process-dependent matrix elements 
$\langle h_1 h_2 |O_i|B \rangle^T $.   
Their phases interfere with the CKM  phase 
contained in some of $\lambda_i$'s.  

One can in principle avoid a direct calculation of 
separate matrix elements and 
employ isospin and $SU(3)_{fl}$ symmetries. The
combinations of the above hadronic matrix elements 
form a smaller set of 
independent flavour-symmetry amplitudes. The latter 
are then  fitted from  the measured experimental observables  
and used to predict other observables 
(see e.g., \cite{Burasetal} and the talk by F.~Schwab). 
Amplitude  decompositions and fits, being useful 
phenomenological tools, provide very 
limited information on  
the details of the quark-gluon dynamics in hadronic  $B$ decays. 
To clarify the role of emission, penguin and annihilation topologies 
and to assess the magnitude of $ SU(3)_{fl}$-symmetry violation 
in charmless channels, 
one has to apply  calculational methods based on QCD.  

Amplitudes with two hadrons in the final state are not directly 
accessible in lattice QCD. To calculate the hadronic 
matrix elements  for $B\to h_1 h_2$, one employs  
approximations and  effective theories valid in 
$m_b\to \infty $  ($E_{h1,h2}\sim m_b/2 \to \infty $) limits. 
The $1/m_b$ ($1/E_h$) expansion is combined with 
the perturbation theory in $\alpha_s$. 
The QCD factorization (QCDF) approach \cite{BBNS}, 
the framework of SCET \cite{SCET}, or a combination of both
(see e.g., \cite{Beneke})  are used. 
In these approaches the input includes heavy-to-light form factors 
(e.g., the $B\to \pi$ form factor at small momentum transfer 
$f_{ B\pi}^+(0)$), as well as  
the light-cone distribution amplitudes (DA's) of $h_1, h_2$ 
and the inverse moment $\lambda_B$ of the B-meson DA.
For example, in the original QCDF analysis of 
$B\to K\pi , \pi\pi$ decays \cite{BBNS} the following parameters 
were adopted:
\begin{equation}
f_{ B\pi}^+(0)=0.28\pm 0.05, ~\lambda_B=350\pm 150 
~\mbox{MeV} \,, 
\label{inputBpi}
\end{equation}
in agreement with the QCD sum rule predictions (see the next section). 
A different approach is PQCD \cite{PQCD} 
where the form factors are assumed to have a perturbative expansion  and the meson wave functions depend on 
transverse momenta. 

Successful phenomenological applications of QCDF or SCET 
to $B\to h_1 h_2$ are possible if all $1/m_b$ effects are under control.
In fact, in QCDF the annihilation contribution to $B\to \pi \pi$, 
as well as the twist-3 part of the spectator-scattering (part of the 
emission topology) contain end-point singularities.
In the ``default version'' of QCDF \cite{BBNS} 
these contributions are  parameterized  
with complex amplitudes, in other words, by additional input
parameters. The latter are expected to be reasonably small
with respect to the leading factorizable amplitude. 
Meanwhile, the current $B\to \pi\pi$ data,
in particular, the unexpectedly large BR($B^0\to\pi^0\pi^0$) 
and direct CP-asymmetry in $B^0\to\pi^+\pi^-$ (observed 
by Belle), cannot be reproduced without inflating at least 
one of the hadronic matrix elements (together with its phase) 
in the decomposition  (\ref{inputBpi}). The uncalculable $O(1/m_b)$ 
complex amplitudes 
mentioned above  offer one possibility  which, however would mean a 
rather poor $1/m_b$ expansion. Another possibility \cite{Beneke} 
is that the NLO, $O(\alpha_s^2)$ spectator-scattering effects 
(not suppressed at $m_b\to \infty$) are large. In this case, 
a better agreement with the 
$B\to\pi \pi $ data is achieved, if both the form factor and inverse moment
in (\ref{inputBpi})  become numerically close to their lower limits.

A different point of view is adopted in the recent phenomenological 
analyses based on SCET \cite{SCET},
where the charming penguin contribution is allowed to vary
(resembling the ansatz suggested earlier in \cite{charmpeng}).
The fit to the $B\to\pi\pi$ data in the SCET approach yields 
a large strong phase for the penguin amplitude  
and a  form factor smaller than in (\ref{inputBpi}): 
$f_{ B\pi}^+(0)=(0.19 \pm 0.01\pm 0.03)(4.25\times 10^{-3}/|V_{ub}|)$.   
In addition, according to \cite{ArnesenLigeti}
(see also the talk by Z.~Ligeti), the annihilation contribution 
in SCET is finite and real.

Summarizing this brief overview, so far there is   
no consensus on the hadronic input in charmless $B$ decays. 
The methods based 
on $1/m_b$ expansion are not in a position to reproduce  
all $B\to\pi\pi$ amplitudes in  agreement with  the current data,
a situation sometimes  called the ``$B\to\pi\pi$ puzzle''.
Only if some of the hadronic matrix elements (leading or suppressed 
in $1/m_b$) are varied as free parameters, 
successful fits to $B\to\pi\pi$ and, eventually also to 
$B\to K\pi$ data are achieved. (A more detailed analysis can be found, 
e.g., in \cite{FeldHurth}.) 

In what follows, I will focus
on a few important questions arising from the above 
discussion: Are the values  of $f_{B\pi}(0)$ and $\lambda_B$  
smaller than  expected earlier?  
How large are  the $SU(3)_{fl}$ violation effects 
in the form factors? 
Is it possible to calculate the hadronic matrix elements
$\langle h_1 h_2 |O_i|B \rangle^T $ at finite $m_b$ 
and assess the  role of various quark topologies 
(emission, penguin, annihilation)?   
I will address these questions using the results obtained from 
QCD sum rules and LCSR.

\section{USE OF QCD SUM RULES AND LCSR}

\subsection{The $B$ meson decay constant}

The decay constant  $f_B$ has no direct relation
to charmless $B$ decays, but plays an important role 
for the normalization of other hadronic matrix elements,
The recent first measurement of $BR(B\to \tau \nu_\tau)$ 
determining $|V_{ub}|f_B$, makes the size of this constant
an even more topical issue. 
The QCD sum rule for $f_B^2$ is obtained from 
the vacuum correlation function of two  heavy-light currents
calculated with the local (condensate) operator-product expansion (OPE) and,  
primarily, at a finite $b$-quark mass. Importantly,
it is also possible to derive the sum rule in HQET 
for the ``static'' constant $\hat{f}_B\simeq f_B \sqrt{m_b}$.
The updated sum rule determination of $f_B$ 
with $O(\alpha_s^2)$ accuracy can be found in the two independent
analyses \cite{JL} and \cite{PenSt}, 
predicting $f_B=210\pm 19$ MeV, $f_{B_s}=244\pm 21$ MeV  
and  $f_B=206\pm 20$ MeV, respectively.  
To compare, the recent lattice $(n_f=3)$ result is
\cite{Okamoto} $f_B=216 \pm 9 \pm 19\pm 7$ MeV,
$f_{B_s}=260 \pm 7 \pm 26\pm 9$ MeV
(see also the talk by V.~Lubicz). 
The first measurement of $BR(B\to \tau \nu_\tau)$ reported by 
Belle Collaboration \cite{BellefB} yields 
$|V_{ub}|f_B= \left(7.73^{+1.24+0.66}_{-1.02 -0.58}\right)
\times 10^{-4}$ GeV. Using the value  
$|V_{ub}|=(4.39\pm 0.19\pm 0.27)\times 10^{-3}$ \cite{HFAG}  
based on the inclusive $b\to u$  analysis,  
one obtains $f_B= 176^{+28+20}_{-23-19}$ MeV. 
There is an agreement with the above QCD predictions 
within the experimental errors and theoretical uncertainties. 
Note, however, that if 
one uses the central value of the experimental result and the 
calculated $f_B$, a smaller central value for $|V_{ub}|$ is obtained.

\subsection{Heavy-to-light Form Factors}

The LCSR for the $B\to \pi$ form factor $f_{B\pi}^+(q^2)$
is derived from the vacuum-pion correlation function
expanded near the light-cone in terms of the pion DA's.
The latter are determined from the two-point QCD sum rules and/or 
from the LCSR for the pion e.m. form factors. For consistency,  
the value of $f_B$ and the quark-hadron duality 
threshold $s_0^B$  are taken from the sum rule 
discussed in the previous subsection.
The recent LCSR result \cite{BZpseud} 
\begin{equation}
f_{B\pi}^+(0)=0.258 \pm 0.031 
\label{fplBZ}
\end{equation}
was obtained with the NLO, $O(\alpha_s)$ (in the twist 2,3 parts) and LO (in
the suppressed twist-4 part) accuracy. 
From this result and the HFAG average 
of $BR(B\to \pi l \nu_l)$ at $0<q^2< 16$ GeV$^2$ the value  
$|V_{ub}|= 3.25\pm (0.17)_{exp}\left(^{+0.54}_{-0.36}\right)_{th}\times
10^{-3}$ is extracted \cite{HFAG}. It is tempting to say that 
this value is smaller than the ``inclusive'' one, and 
closer to $|V_{ub}|$ obtained from the central value of 
$BR(B\to \tau\nu_\tau)$ combined with  the QCD prediction for $f_B$.   
However, the errors and uncertainties are still too large 
for a definitive comparison.
A value of the $B\to\pi$ form factor
\begin{equation}
f^+_{B\pi}(0)=0.26\pm 0.02 \pm 0.03\,,
\label{fpl}
\end{equation}
close to (\ref{fplBZ}) was reproduced from the same LCSR  in \cite{BpipiA} 
(without the small twist-3 $O(\alpha_s)$ correction). 
Here the uncertainties induced by the pion DA's and 
by other sum rule parameters are shown, respectively.

The $B\to\pi$ form factor is accessible on the lattice only 
at sufficiently large $q^2$. It is possible to use 
dispersion bounds to extrapolate the lattice QCD 
results to $q^2=0$. The two recent analyses  yield 
$f^+_{B\pi}(0)=0.25\pm 0.06$ (the version 
without the SCET point) \cite{AGRS} and $f^+_{B\pi}(0)=0.25\pm 0.04$
~\cite{BecherHill}, in a good agreement with the above LCSR results. 
The interval of the form factor advocated by the SCET 
fits to $B\to\pi\pi$ still agrees with both LCSR and lattice QCD,  but only within 
uncertainties.  

Combining the  average $BR(B\to \pi l\nu_l)= (1.34\pm 0.08\pm 0.08)\times 10^{-4}$ 
\cite{HFAG}, the slope parameter $\alpha_{B\pi}= 0.61\pm 0.09$  fitted 
from the $q^2$ distribution in $B\to \pi l\nu_l$
by BaBar Collaboration \cite{Babar}, 
and the new Belle measurement \cite{BellefB} 
$BR(B\to \tau \nu_\tau)=(1.06^{ +0.34+0.18}_{ -0.28-0.16})\times 10^{-4}$
with the interval of QCD 
predictions $f_B=210\pm 20$ MeV, 
it is possible to calculate the $B\to \pi$ form factor at zero 
momentum transfer, independent of $|V_{ub}|$. One obtains: 
\begin{equation}
f_{B\pi}^+(0)=0.24 \pm 0.01\pm 0.01\pm 0.04 \pm 0.02\,,
\end{equation}
where the first three errors are experimental (semileptonic width, shape,
leptonic width) and the fourth error
is due  to the theoretical uncertainty of $f_B$.
The central value of this estimate nicely agrees with the LCSR prediction 
but  the errors are again too large to exclude a smaller form factor.

\subsection{Inverse moment of $B$-meson DA}

Recently, a new type of LCSR was derived from
the correlation function between the vacuum and B meson, 
relating the $B\to\pi$ 
form factor at small momentum transfer to the $B$-meson
DA's \cite{KMO}. A similar approach was suggested in the framework of
SCET  in  \cite{DeFFH}. In the leading order, this sum rule establishes 
a simple relation between the combination  
$(m_B\lambda_B f^+_{B\pi}(0)/f_B)$  and the parameters in the pion 
channel. (Note that in the above combination the heavy-quark mass 
dependence drops out in the $m_b\to\infty$ limit). 
Using the results for $f^+_{B\pi}(0)$ and $f_B$ 
from the sum rules described in the subsections 3.1 and 3.2,
respectively, a new estimate for the inverse moment 
$\lambda_B=460 \pm 160$ MeV 
was obtained, in agreement with the HQET sum rule estimates  
\cite{GN_BIK} and with the interval
for  $\lambda_B$ in (\ref{inputBpi}).
Note that the product of the form factor and 
inverse moment is fixed by the new sum rule.
The latter is therefore violated if both $f^+_{B\pi}(0)$
and $\lambda_B$ decrease e.g., up to their lower limits in 
(\ref{inputBpi}). A more detailed numerical analysis of this relation 
including the power suppressed corrections is currently in progress.   

\subsection{$SU(3)_{fl}$ violation  in $B$ decays }

QCD sum rules allow to calculate the differences between the 
hadronic matrix elements with kaons and pions in terms of $m_s\neq m_{u,d}$
and the ratios of 
the strange- and nonstrange quark condensates. In this way, the 
ratio  $f_K/f_\pi$ is reproduced and 
the $SU(3)_{fl}$-asymmetries in the kaon DA's vs pion DA's 
are predicted (see, e.g. \cite{KMMa1K}
and the recent comprehensive analysis in \cite{BBL}).
After including the {\em calculated } $SU(3)_{fl}$-violation effects 
into LCSR, one predicts the ratios of the form factors with and 
without strange hadrons. 
The results obtained in \cite{KMMsu3}  have been 
updated in \cite{KMMa1K} (after an important  change of the first 
Gegenbauer moment in the twist-2 kaon DA) to
the following intervals:
\begin{eqnarray}
  \frac{f^+_{BK}(0)}{f^+_{B\pi}(0)}
    = 1.36^{+0.12}_{-0.09}\,, ~~
  \frac{f^+_{B_s K}(0)}{f^+_{B\pi}(0)}
    = 1.21^{+0.14}_{-0.11}\,.
\end{eqnarray}
The  $SU(3)_{fl}$-violation effects are large, and they influence 
the flavour-symmetry relations for nonleptonic charmless $B$-decays. 
The prediction for $BR(B_s\to K^+K^-)$ obtained in \cite{Burasetal} on the 
basis of these relations, including 
the above sum rule  predictions for the form factor ratios, 
have recently been confirmed by CDF Collaboration \cite{CDFBs}.

\subsection{Hadronic matrix elements for $B\to \pi\pi$}

An extension of LCSR technique \cite{AKBpipi} allows to 
calculate various hadronic matrix elements
$\langle h_1 h_2 |O_i|B \rangle^T $
and compare them with each other. An important 
study case is $\bar{B}^0\to \pi^+\pi^-$.   
One starts with the correlation function: 
\begin{eqnarray}
&F^{(O_i)}_\alpha(p,q,k)=-\int d^4x~e^{-i(p-q)x}\int d^4y~e^{i(p-k)y} 
\nonumber\\
&\times
\langle 0|T\left\{j_{\alpha 5}^{(\pi)}(y) O_i(0)
j_5^{(B)}(x)\right\}|\pi^-(q)\rangle,
\label{eq-F}
\end{eqnarray}
where $j_{\alpha 5}^{(\pi)}=\bar{u}\gamma_\alpha\gamma_5 d$ and
$j_5^{(B)}=m_b \bar{b}i\gamma_5 d$
are the quark currents 
interpolating the pion and the $B$ meson,
respectively. The correlation function $F^{(O_i)}$ is 
calculated in QCD at large 
spacelike external momenta squared
$(p-k)^2,(p-q)^2, P^2=(p-q-k)^2$, in a form of 
OPE with the pion DA's, that is, using the same long-distance
input, as  in the LCSR for the $B\to \pi$ form factor
(see sect. 3.2). Furthermore, both hard- and soft-gluon 
effects are included, contributing to different terms of 
the OPE. The result for the correlation function is then 
matched to the hadronic dispersion relations, subsequently
in the pion ($j_{\alpha 5}^{(\pi)}$) and $B$ meson 
($j_5^{(B)}$) channels. The procedure is 
formulated in such a way 
that the final sum rule relation contains the desired 
matrix element $\langle\pi^+\pi^- | O_i|\bar{B}^0 \rangle$ . 
The calculation is fulfilled
at finite $m_b$. A further expansion of the sum rule in $1/m_b$ 
allows to compare the finite $m_b$ results for the hadronic 
matrix elements with the corresponding QCDF predictions, 
and, moreover, to estimate the uncalculable $1/m_b$ corrections.
For a given operator $O_i$ in the correlation function
(\ref{eq-F}) various contractions of quark fields are possible. 
Collecting the lowest contributions to OPE, one 
identifies the diagrams with emission, penguin and 
annihilation topologies. If one retains only diagrams 
with a topology $T$ in the OPE, the sum rule result 
is interpreted as $\langle\pi^+\pi^- | O_i|\bar{B}^0 \rangle_T$.
Note that an important step in obtaining the LCSR results
for $B\to\pi\pi$ hadronic matrix elements
is the analytical continuation 
from a large spacelike $P^2<0$ ($|P^2|\gg \Lambda_{QCD}^2$) 
to the large timelike  $P^2=m_B^2$. 
The imaginary part 
(discontinuity) in $P^2$ generated by this continuation
is identified with the strong phase of the hadronic
matrix element in the (local) quark-hadron duality approximation.  
This duality assumption introduces an 
additional  ``systematic'' uncertainty.

The details of the procedure and calculation for various 
topologies can be found in \cite{AKBpipi,LCSRBpipi,BpipiA}. 
In particular, a  finite result for the 
hadronic matrix element $\langle\pi^+\pi^- | O_1^u|\bar{B}^0 \rangle_A$ of 
the current-current operator 
$O_1^u=(\bar{d}\Gamma_\mu u)(\bar{u}\Gamma^\mu b)$  is obtained, 
with an imaginary part which contributes to 
the strong phase. The origin of this annihilation phase
at the diagram level is explained in \cite{BpipiA}. 
In addition, an important factorizable annihilation 
contribution from the quark-penguin operator $O_6$ 
is found. For the contributions with 
hard gluons  the method suggested in \cite{AKBpipi} was modified 
in \cite{BpipiA},  to avoid complicated two-loop multi-scale
diagrams. Instead of performing the QCD calculation of 
the vacuum-to-pion correlation function (\ref{eq-F}),
one starts from  the pion-pion correlator, thereby reducing the 
calculation to one-loop diagrams.
\newcommand{\GeV}{~\textrm{GeV}}
\newcommand{\MeV}{~\textrm{MeV}}
The input parameters used in the LCSR 
for $B\to\pi\pi$ hadronic matrix elements
are the same as in the LCSR for $f_{B\pi}^+$.
This form factor determines the factorizable  
$B\to \pi\pi$ amplitude, that is, the hadronic matrix element
of the $O_1^u$   in the emission topology:
\begin{equation} 
\langle \pi^+\pi^-| O_1^u|\bar{B}^0 \rangle_E\simeq if_\pi
f^+_{B\pi}(0)m_B^2 \,. 
\label{Afact}
\end{equation}
The LCSR obtained in  \cite{AKBpipi,LCSRBpipi,BpipiA}
were used to estimate the hadronic matrix elements
$\langle\pi^+\pi^- | O_i|\bar{B}^0 \rangle_T$
of all effective operators $O_i$ (except the electroweak penguin operators) 
with nonfactorizable emission, penguin and annihilation topologies. 

The ratio of the  annihilation matrix element 
to the factorizable amplitude (\ref{Afact}), defined as   
\begin{equation}
r_A^{(\pi\pi)}=\frac{\langle\pi^+\pi^-|O_1^u|\bar{B}^0 \rangle_A }{
2\langle \pi^+\pi^-| O_1^u|\bar{B}^0 \rangle_E }\,,
\label{ratioA}
\end{equation} 
as well as the analogous ratios for other 
topologies and operators
parameterize the nonfactorizable effects.
The numerical results are :

$*$ nonfactorizable emission
\begin{eqnarray}
r_E^{(\pi\pi)}=\Big[\left(1.8^{+0.5}_{-0.7}\right)\times
10^{-2}\Big]_\textrm{soft}
\nonumber\\
+\Big[\Big(-1.9^{+0.5}_{-0.1}+
i\left(-3.6^{+1.0}_{-0.4}\right)\Big)\times 10^{-2}\Big]_\textrm{hard}\,,
\end{eqnarray}
where the contribution of the hard gluon is a preliminary
result (with twist 2 accuracy), 
 
$*$ charming penguin:
\begin{equation}
r_{P_c}^{(\pi\pi)}=\left[-0.18^{+0.06}_{-0.68}
+i\left(-0.80^{+0.17}_{-0.08}\right)\right]\times 10^{-2}\,,
\nonumber
\end{equation}

$*$ annihilation:
\begin{equation}
r_A^{(\pi\pi)}
= \left[-0.67^{+0.47}_{-0.87}+
i\left(3.6^{+0.5}_{-1.1}\right)\right]\times 10^{-3}\,.
\nonumber
\end{equation}

These results are encouraging for the method because all nonfactorizable 
effects including their imaginary parts (contributions to the  
strong phase)  are found small, so that the 
light-cone OPE  can be trusted. 
Moreover, expanding LCSR in $1/m_b$, one reproduces  
the hierarchy of QCDF, with a possibility  to estimate also 
the $1/m_b$ effects (e.g. the nonfactorizable soft-gluon contributions
in the emission topology \cite{AKBpipi}). 

At the same time, the results are rather discouraging 
for the phenomenology, because sum rules 
do not reveal any ``hidden'' large effect in 
penguin and/or  annihilation topologies, 
thereby supporting the default version of QCDF.
Adding up all small effects estimated with LCSR 
one obtains (for the CKM angle $\gamma=(58.5\pm 10)^o$)):
\begin{eqnarray}
BR(B^+ \to \pi^+ \pi^0)&=&\left(6.6^{+1.8+0.8}_{-1.3-0.8}
\right)\times \!10^{-6} \nonumber \\ 
BR(B^0 \to \pi^+ \pi^-)&=& 
\left(9.7^{+2.3+1.2}_{-1.9-1.2}
\right)\times \!10^{-6} \nonumber \\
BR(B^0 \to \pi^0 \pi^0)&=& 
\left(0.25^{+0.12+0.07}_{-0.08-0.06}
\right)\times \!10^{-6}\,.\nonumber
\end{eqnarray}
The last two predictions significantly differ from the current 
experimental averages 
\cite{HFAG} for $BR(B^0 \to \pi^+ \pi^-)$ and $BR(B^0 \to \pi^0 \pi^0)$. 
For brevity I do not show the LCSR results for $CP$ asymmetries 
presented in \cite{BpipiA}, where a very small direct asymmetry
for $B^0 \to \pi^+ \pi^-$ is predicted.

The comparison with the data based on the most general isospin
expansion of the decay amplitudes \cite{BpipiA}
reveals that the isospin-two amplitude  
determining the $B^- \to \pi^- \pi^0$ decay 
is in a reasonable agreement with theoretical 
predictions, (using the 
$B\to \pi$ form factor calculated 
from LCSR), whereas one needs 
additional contributions to the isospin-zero amplitude  
generated by the $\Delta I = 1/2$ 
pieces of the effective Hamiltonian.

\section{CONCLUSIONS}

QCD sum rules for $f_B$ and LCSR for 
$B\to\pi$ form factor agree with the lattice QCD results, and 
hint at a smaller value of $|V_{ub}|$ than the ``inclusive'' one,
but the uncertainties remain too large for a decisive comparison.
A scenario with both small $f_{B\pi}^+(0)$ and $\lambda_B$ 
is  disfavored by the new sum rule for the product of these two
parameters. LCSR calculations of hadronic matrix elements 
$\langle \pi\pi | O_i | B\rangle_T$ reveal 
suppressed nonfactorizable effects with small strong phases in $B\to \pi\pi$ 
(probably also in $B\to K\pi$, a more detailed analysis 
including $SU(3)_{fl}$-violation effects is in progress).
Thus, the origin of an additional isospin-zero amplitude 
and related large strong phase indicated by the $B\to \pi\pi$
data \footnote{see e.g., the talk by T.N.~Pham and \cite{KaidVys}} cannot be identified.\\

I thank  Giulia Ricciardi and other organizers
for a very interesting and enjoyable workshop. 
I gratefully acknowledge collaboration with Thomas Mannel, 
 Blazenka Meli\'c, Martin Melcher and Nils Offen on the topics 
discussed in this talk. This work is supported
by Deutsche Forschungsgemeinschaft (DFG) under the project
DFG KH205/1-1.


\begin{thebibliography}{9}

\bibitem{HFAG} 
    [Heavy Flavor Averaging Group (HFAG)],
  arXiv:hep-ex/0603003.



\bibitem{Burasetal}
  A.~J.~Buras, R.~Fleischer, S.~Recksiegel and F.~Schwab,
  Nucl.\ Phys.\ B {\bf 697} (2004) 133.


\bibitem{BBNS}
M.~Beneke, G.~Buchalla, M.~Neubert and C.~T.~Sachrajda,
Phys.\ Rev.\ Lett.\  {\bf 83} (1999) 1914;
Nucl.\ Phys.\ B {\bf 606} (2001) 245\,.

\bibitem{SCET}
  C.~W.~Bauer, D.~Pirjol, I.~Z.~Rothstein and I.~W.~Stewart,
  Phys.\ Rev.\ D {\bf 70} (2004) 054015;
  C.~W.~Bauer, I.~Z.~Rothstein and I.~W.~Stewart,
  arXiv:hep-ph/0510241;\\
  J.~Chay and C.~Kim,
  Nucl.\ Phys.\ B {\bf 680} (2004) 302.

\bibitem{Beneke}
  M.~Beneke and S.~J\"ager,
  arXiv:hep-ph/0512351.


\bibitem{PQCD} 
  Y.~Y.~Keum and H.~n.~Li,
  Phys.\ Rev.\ D {\bf 63} (2001) 074006;
  Y.~Y.~Keum, H.~n.~Li and A.~I.~Sanda,
  Phys.\ Lett.\ B {\bf 504} (2001) 6;
  H.~n.~Li and S.~Mishima,
  Phys.\ Rev.\ D {\bf 73} (2006) 114014.


\bibitem{charmpeng}
  M.~Ciuchini, E.~Franco, G.~Martinelli and L.~Silvestrini,
  Nucl.\ Phys.\ B {\bf 501}, 271 (1997);
  M.~Ciuchini, E.~Franco, G.~Martinelli, A.~Masiero, M.~Pierini and L.~Silvestrini,
  arXiv:hep-ph/0407073.

\bibitem{ArnesenLigeti}
  C.~M.~Arnesen, Z.~Ligeti, I.~Z.~Rothstein and I.~W.~Stewart,
  arXiv:hep-ph/0607001.

\bibitem{FeldHurth}
  T.~Feldmann and T.~Hurth,
  JHEP {\bf 0411} (2004) 037.

\bibitem{JL}
  M.~Jamin and B.~O.~Lange,
  Phys.\ Rev.\ D {\bf 65} (2002) 056005.

\bibitem{PenSt}
  A.~A.~Penin and M.~Steinhauser,
  Phys.\ Rev.\ D {\bf 65} (2002) 054006.

\bibitem{Okamoto}
M.~Okamoto,
PoS {\bf LAT2005} (2006) 013
[arXiv:hep-lat/0510113].



\bibitem{BellefB}
K.~Ikado {\it et al.} [Belle Collaboration]
arXiv:hep-ex/0604018.


\bibitem{BZpseud}
  P.~Ball and R.~Zwicky,
  Phys.\ Rev.\ D {\bf 71} (2005) 014015\,;
  hep-ph/0507076\,.
 

\bibitem{BpipiA}
  A.~Khodjamirian, T.~Mannel, M.~Melcher and B.~Meli\'c,
  Phys.\ Rev.\ D {\bf 72} (2005) 094012.



\bibitem{AGRS}
M.~C.~Arnesen, B.~Grinstein, I.~Z.~Rothstein and I.~W.~Stewart,
Phys.\ Rev.\ Lett.\  {\bf 95} (2005) 071802.

\bibitem{BecherHill}
T.~Becher and R.~J.~Hill,
Phys.\ Lett.\ B {\bf 633}, 61 (2006).


\bibitem{Babar}
  B.~Aubert {\it et al.}  [BABAR Collaboration],
  Phys.\ Rev.\ D {\bf 72} (2005) 051102


\bibitem{KMO}
  A.~Khodjamirian, T.~Mannel and N.~Offen,
  Phys.\ Lett.\ B {\bf 620} (2005) 52\,.


\bibitem{DeFFH}
  F.~De Fazio, T.~Feldmann and T.~Hurth,
  Nucl.\ Phys.\ B {\bf 733} (2006) 1.




\bibitem{GN_BIK}
  A.~G.~Grozin and M.~Neubert,
  Phys.\ Rev.\ D {\bf 55} (1997) 272\,;
  V.~M.~Braun, D.~Y.~Ivanov and G.~P.~Korchemsky,
  Phys.\ Rev.\ D {\bf 69}, 034014 (2004).


\bibitem{KMMa1K}
A.~Khodjamirian, T.~Mannel and M.~Melcher,
Phys.\ Rev.\ D {\bf 70}, 094002 (2004)

\bibitem{BBL}
  P.~Ball, V.~M.~Braun and A.~Lenz,
  JHEP {\bf 0605}, 004 (2006)
  [arXiv:hep-ph/0603063].


\bibitem{KMMsu3}
  A.~Khodjamirian, T.~Mannel and M.~Melcher,
  Phys.\ Rev.\ D {\bf 68} (2003) 114007.

\bibitem{CDFBs}
  A.~Abulencia  [CDF Collaboration],
  arXiv:hep-ex/0607021.


\bibitem{AKBpipi}
A.~Khodjamirian,
Nucl.\ Phys.\ B {\bf 605} (2001) 558\,.

\bibitem{LCSRBpipi}
A.~Khodjamirian, T.~Mannel and P.~Urban,
Phys.\ Rev.\ D {\bf 67} (2003) 054027;
  A.~Khodjamirian, T.~Mannel and B.~Meli\'c,
  Phys.\ Lett.\ B {\bf 571} (2003) 75\,.




\bibitem{KaidVys}
  A.~B.~Kaidalov and M.~I.~Vysotsky,
  arXiv:hep-ph/0603013.






\end{thebibliography}
\end{document}